\documentclass[prl,twocolumn,preprintnumbers,showpacs]{revtex4-1}

\usepackage[dvips]{color,graphicx}
\usepackage{bm}
\usepackage{amsmath,amssymb}

\newcommand{\bB}{\boldsymbol{B}}
\newcommand{\bA}{\boldsymbol{\mathcal{A}}}
\newcommand{\bj}{\boldsymbol{j}}
\newcommand{\calA}{\mathcal{A}}
\newcommand{\calC}{\mathcal{C}}
\newcommand{\calD}{\mathcal{D}}

\newcommand{\GeV}{\;\text{GeV}}
\newcommand{\kmax}{k_{\rm max}}

\begin{document}

\title{Dielectric Correction to the Chiral Magnetic Effect}

\author{Kenji Fukushima} 
\author{Marco Ruggieri}  
\affiliation{Yukawa Institute for Theoretical Physics,
 Kyoto University, Kyoto 606-8502, Japan}

\begin{abstract}
 We derive an electric current density $\bj_{\rm em}$ in the presence
 of a magnetic field $\bB$ and a chiral chemical potential $\mu_5$.
 We show that $\bj_{\rm em}$ has not only the anomaly-induced term
 $\propto \mu_5 \bB$ (i.e.\ Chiral Magnetic Effect) but also a
 non-anomalous correction which comes from interaction effects and
 expressed in terms of the susceptibility.  We find the correction
 characteristically dependent on the number of quark flavors.  The
 numerically estimated correction turns out to be a minor effect on
 heavy-ion collisions but can be tested by the lattice QCD
 simulation.
\end{abstract}

\preprint{YITP-10-31}
\maketitle


The Chiral Magnetic Effect
(CME)~\cite{Kharzeev:2007jp,Fukushima:2008xe} is a general mechanism
to induce an electric current $\bj_{\rm em}$ along the direction of an
external magnetic field $\bB$ for systems that have a non-vanishing
chirality charge $N_5=N_R-N_L$.  Using the chiral chemical potential
$\mu_5$ to express $N_5\neq0$ in the grand canonical ensemble one can
find the electric current density given simply
as~\cite{Fukushima:2008xe}
\begin{equation}
 \bj_{\rm em} = \frac{e^2 \mu_5 \bB}{2\pi^2} \;,
\label{eq:CMEcurrent}
\end{equation}
which originates from the quantum anomaly and thus is an exact
relation insensitive to any infrared scales such as the particle mass,
temperature, etc.  This type of anomaly relation is quite generic and
also discussed in various
contexts~\cite{Alekseev:1998ds,Metlitski:2005pr,Newman:2005as,Gorbar:2009bm}.
Particularly, in prior to the CME, a dual situation had attracted
attention; it is the axial current $\bj_5$ that is induced by $\bB$ if
the quark chemical potential $\mu_q$ is non-zero, which is parallel to
each other under the replacement $\mu_5\leftrightarrow\mu_q$ and
$\bj\leftrightarrow\bj_5$.  The resulting axial current may be
realistic in the cores of compact stellar
objects~\cite{Metlitski:2005pr} or in some condensed matter
systems~\cite{Gorbar:2009bm}.  Generally speaking, the structure of
Eq.~\eqref{eq:CMEcurrent} can also arise from low-energy effective
descriptions.  That is, from symmetry reason, the anomalous current is
to be expressed as
$j^\mu_{\rm em} = -(\partial_\nu \phi)\tilde{F}^{\mu\nu}$ where $\phi$
is a pseudo-scalar field \cite{Asakawa:2010bu}.  Under the magnetic
field $B^i=\tilde{F}^{0i}$ the current is thus proportional to $\bB$
and $\partial_0\phi$ that translates into $\mu_5$ of
Eq.~\eqref{eq:CMEcurrent}.  This gives a physical interpretation of
$\mu_5$ as a time derivative of the pseudo-scalar condensate.
Equivalently one can also say that $\mu_5$ appears as a result of a
time derivative on the strong $\theta$-angle
parameter~\cite{Fukushima:2008xe} which leads to pseudo-scalar meson
enhancement through mixture of the $\sigma$ and $\eta_0$
condensates~\cite{Kharzeev:1998kz,Mizher:2008dj}.

It is tremendously important to verify Eq.~\eqref{eq:CMEcurrent} in
order to quantify the CME in such a way that theoretical predictions
can be compared to experimental
measurements~\cite{Voloshin:2004vk,STAR}.  At the same time the
estimate of background (non-topological)
effects~\cite{Wang:2009kd,Fukushima:2009ft,Millo:2009ar,Bzdak:2009fc}
is indispensable for experimental confirmation of the CME.\ \  The
purpose of this Letter is to point out that the induced current
\eqref{eq:CMEcurrent} itself must receive a non-anomalous correction
from interaction effects even at the level of the mean-field
quasi-particle description.  The notable feature of the correction is
that the coefficient in Eq.~\eqref{eq:CMEcurrent}, which seems to be
protected by anomaly, is modified by a factor.  Existence of such
correction is a novel insight in theory.


In this Letter we do not directly consider the interaction mediated by
gauge bosons but instead make use of an effective form of the
interaction in terms of fermionic degrees of freedom.  It should be a
reasonable approximation to utilize the current-current interaction,
\begin{equation}
 \mathcal{L}_V = -G_V (\bar{\psi}\gamma_\mu \psi)
  (\bar{\psi}\gamma^\mu \psi) \;,
\label{eq:vector}
\end{equation}
as long as the typical interaction energy is lower than the gauge
boson mass, which is the case for Fermi's effective theory of the weak
interactions.  In QCD the one-gluon exchange can result in an
interaction form of the color current in which the color generators
are inserted.  It is then possible to extract a specific
form~\eqref{eq:vector} out from the Fierz transformation.  Accordingly
$G_V$ should be of order $\sim g^2/M_g^2$ where $g$ is the gauge
coupling constant and $M_g$ is the gauge boson mass.  [Gluons should
  be massive non-perturbatively though their mass is zero at the QCD
  Lagrangian level.]  For the moment we treat $G_V$ as just a
parameter and will plug a concrete value in later discussions.

The vector interaction~\eqref{eq:vector} is special in QCD;  it is
invariant under chiral rotations and thus its presence is naturally
anticipated.  Introducing a mean-field
$j^\mu=\langle\bar{\psi}\gamma^\mu\psi\rangle$ the
interaction~\eqref{eq:vector} is decomposed into
$\mathcal{L}_V\to -G_V j^{z2}+2G_V j^z \bar{\psi}\gamma^3\psi$, then,
apart from quantum gauge fluctuations, we have the kinetic term;
$\mathcal{L}_{\rm kin} = \bar{\psi} \bigl(i\gamma_\mu \calD^\mu
- M + \mu_5 \gamma^0\gamma^5 \bigr)\psi$, where $M$ is a mass which
may be dynamically generated, but for the present purpose the
microscopic origin of $M$ is irrelevant.  In writing the above we have
introduced an effective (classical) gauge field in
$\calD^\mu=\partial^\mu - i\calA^\mu$ given by
\begin{equation}
 \calA^x = \calA^y = 0 \;,\quad \calA^z = -2G_V j^z \;,
\label{eq:A}
\end{equation}
for which the coupling constant is chosen as unity.  [This is a
  convenient choice for later generalization to QCD with quark flavors
  having different electric charges.]  The grand potential (divided by
the volume $V$) can be read from the zero-point oscillation energy in
addition to the mean-field condensation energy, i.e.\
\begin{equation}
 \Omega/V =  G_V j^{z2} - \frac{|eB|}{2\pi}
  \sum_{s,k} \alpha_{sk} \int\frac{d p^z}{2\pi} \,\omega_s(p)  \;,
\label{eq:Omega}
\end{equation}
where $s$ is the spin and $k$ refers to the Landau level.  The spin
degeneracy factor is taken care of $\alpha_{sk}$ defined as
\begin{equation}
 \alpha_{sk} = \left\{ \begin{array}{ll}
  \delta_{s,+1} & \text{ for } \quad k=0,\;\; eB>0  \;,\\
  \delta_{s,-1} & \text{ for } \quad k=0,\;\; eB<0  \;,\\
  1            & \text{ for } \quad k\neq0          \;.
 \end{array} \right.
\end{equation}
Now the quasi-particle dispersion relations are derived from the
eigenvalues of the Dirac operator which are~\cite{Fukushima:2008xe};
$\omega_s^2 = M^2 + [ |{\bm p}| + \text{sgn}(p^z) s \mu_5 ]^2$,
where $|\bm p|^2 = (p^z + \calA^z)^2 + 2|eB| k$ with $k$ being a
non-negative integer to label the Landau level.

The stationary condition for $\Omega$ with respect to $\bj$,
i.e.\ $\partial\Omega/\partial\bj=0$, yields a self-consistent
condition to determine the current density,
\begin{equation}
 \frac{\partial\Omega}{\partial\bj}=0 \;\rightarrow\quad
  \bj = \frac{\partial (\Omega/V)}{\partial{\bA}} \;.
\label{eq:current}
\end{equation}
We note that the derivative with respect to $\bA$ acts only on the
latter term of \eqref{eq:Omega} that explicitly depends on $\bA$.

The expression \eqref{eq:current} looks like a standard one utilized
in Ref.~\cite{Fukushima:2008xe} but the essential difference is that
the derivative should be evaluated at non-zero $\bA$ which is
specified by \eqref{eq:A} and thus $\bj$ should be solved in a
self-consistent manner.

By taking the derivative explicitly on Eq.~\eqref{eq:Omega} only the
surface terms contribute to the current.  Here we assume some
regularization scheme and introduce a ultraviolet momentum scale
$\Lambda_k$ in such a way that
$\Lambda_k^2+2|eB|k = \Lambda^2$.  Then the surface terms are picked
up as
\begin{equation}
 \frac{\partial(\Omega/V)}{\partial \calA^z}\biggr|_{\calA}
  = \frac{|eB|}{4\pi^2}\biggl[ \bigl(g_+ - g_-\bigr)
   + \sum_{k=1}^{\kmax} \sum_s
   \bigl( f_{s+} - f_{s-} \bigr) \biggr] \;,
\label{eq:deriv}
\end{equation}
with the energies at $p^z=\pm\Lambda_k$, namely,
\begin{align}
 g_{\pm} &= \sqrt{M^2 + \bigl(|\pm\Lambda_0+\calA^z|
  \pm\text{sgn}(eB)\mu_5 \bigr)^2} \;,\notag\\
 &\approx \Lambda_0 \pm \calA^z \pm \text{sgn}(eB)\mu_5 \;,
\label{eq:g}
\end{align}
for the Landau zero-mode $k=0$ and the second line is an approximation
valid for sufficiently large $\Lambda$.  In the same way we have
\begin{align}
 f_{s\pm} &= \sqrt{M^2 + \bigl( \sqrt{(\pm\Lambda_k + \calA^z)^2
  +2|eB|k} + s\mu_5\bigr)^2 } \;,\notag\\
 &\approx \Lambda \pm \frac{\Lambda_k}{\Lambda}\calA^z \;,
\label{eq:f}
\end{align}
for the Landau non-zero modes $k>0$.  From our definition of
$\Lambda_k$ and $\Lambda$ it is obvious that
$\kmax=\lfloor\Lambda^2/(2|eB|)\rfloor$.

It should be noted that the first contribution in \eqref{eq:deriv}
involving $g_\pm$ leads to the electric current \eqref{eq:CMEcurrent}
in the limit of vanishing $\bA$.  The latter term involving $f_{s\pm}$
is simply zero if $\bA$ is absent.

When $\Lambda$ is sufficiently larger than $|eB|$, the sum over $k$
can be well approximated as
$\sum_k(\Lambda_k/\Lambda)\approx \Lambda^2/(3|eB|)$.  Using this we
reach,
\begin{equation}
 j^z = \frac{|eB|}{2\pi^2}
  \biggl[ \,\text{sgn}(eB)\mu_5 - 2 G_V\biggl(1
  +\frac{2\Lambda^2}{3|eB|}\biggr)j^z\, \biggr] \;.
\label{eq:unrenorm-j}
\end{equation}
From the above one might be able to solve $j$, though the divergent
term still remains.  The expression is not physically meaningful yet
as it is.  One needs to formulate the renormalization procedure that
we address in what follows below.

Here it should be mentioned that our result \eqref{eq:unrenorm-j} is
quite analogous to what is discussed in Ref.~\cite{Gorbar:2009bm}.
The formulation may look different since no current-current
interaction was introduced in Ref.~\cite{Gorbar:2009bm} but familiar
Nambu--Jona-Lasinio (NJL) type (scalar and pseudo-scalar) four-fermion
interactions were considered with the Dyson-Schwinger equation which
goes beyond the present mean-field analysis.  Nevertheless our
treatment suffices to grasp the essential point of
Ref.~\cite{Gorbar:2009bm}.  That is, the interaction term like
Eq.~\eqref{eq:vector} would be always induced by mixing interactions
from scalar and pseudo-scalar channels, which might be the case in the
Dyson-Schwinger calculation.


For the purpose of renormalization let us consider the following
susceptibility defined by
\begin{equation}
 \calC = \frac{\partial^2 (\Omega/V)}{\partial \calA^{z2}}
  \biggr|_{\calA}
  = \frac{|eB|}{2\pi^2}\biggl(1+\frac{2\Lambda^2}{3|eB|}\biggr) \;,
\label{eq:C}
\end{equation}
which is still divergent.  Because this susceptibility can be
interpreted as the gauge boson (screening) mass, it must become
vanishing in the vacuum (i.e.\ for $B=0$) so that the gauge invariance
is maintained.  This imposes the following requirement;
$\calC\rightarrow0$ for $B\rightarrow0$, which sets a natural
condition for minimal subtraction.  In the definition of $\calC$ in
Eq.~\eqref{eq:C} the second term is divergent as $\Lambda^2/(3\pi^2)$,
which must be subtracted through renormalization.  Consequently the
renormalized susceptibility should be given by the finite first term;
\begin{equation}
 \calC_R = \frac{|eB|}{2\pi^2} \;.
\label{eq:Cren}
\end{equation}
This result is consistent with the conclusion in
Ref.~\cite{Fukushima:2009ft} in which an independent derivation of the
susceptibility \eqref{eq:Cren} is given by means of the linear
response formula with the current generation due to anomaly
\cite{Fukushima:2010vw}.  Since the derivation in
Ref.~\cite{Fukushima:2009ft} is free from ultraviolet divergence, this
confirms the validity of Eq.~\eqref{eq:Cren}.

It is straightforward to solve \eqref{eq:unrenorm-j} with respect to
$j^z$ with $\calC$ replaced by $\calC_R$.  We finally find,
\begin{equation}
 \bj = \frac{1}{1+2G_V \calC_R} \cdot
  \frac{\mu_5\,e\bB}{2\pi^2} \;,
\label{eq:jren}
\end{equation}
from which, apart from flavor complication in QCD which will be
discussed later, the electric current density is simply
$\bj_{\rm em}=e\bj$.  If we interpret the coefficient in front of
$\bB$ as the chiral magnetic conductivity~\cite{Kharzeev:2009pj},
Eq.~\eqref{eq:jren} means that the conductivity should be corrected by
the in-medium (i.e.\ magnetic field induced) (generalized) dielectric
constant $\kappa$~\cite{FW}.  We here define $\kappa$ by
\begin{equation}
 \bj_{\rm em} = \kappa \cdot \bj_{\rm em}(G_V=0) \;.
\end{equation}
We note that $\kappa$ deduced from Eq.~\eqref{eq:jren} is a standard
formula representing the screening effects by vacuum
polarization~\cite{FW}.

As we pointed out, the current-current term like Eq.~\eqref{eq:vector}
should be generally present as a result of non-perturbative
interactions, and thus Eq.~\eqref{eq:CMEcurrent} is no longer the
exact answer in the fully interacting case.  In this sense we would
claim that the holographic calculations as in Ref.~\cite{Yee:2009vw}
may miss some back-reaction because the correction
factor~\eqref{eq:jren} has not been found there.  This difference
might explain subtleties on the holographic chiral magnetic current
which are still under dispute \cite{Yee:2009vw,Rebhan:2009vc}.


Let us generalize our result \eqref{eq:jren} to the QCD problem.
Then, the complication comes from flavor degrees of freedom.  In what
follows we will address the one-flavor, two-flavor, and three-flavor
cases in order.

\paragraph{One-flavor case:}
For the case with only one flavor the modification is only the color
factor $N_c=3$ and also that $e$ may be replaced by the quark electric
charge $q$.  Then, the dielectric constant is almost trivially,
\begin{equation}
 \kappa = \biggl( 1 + \frac{3}{\pi^2}G_V |qB| \biggr)^{-1} \;,
\label{eq:one}
\end{equation}
which is unity at $G_V=0$ and goes to zero as $G_V\to\infty$.

\begin{figure}
 \includegraphics[width=0.85\columnwidth]{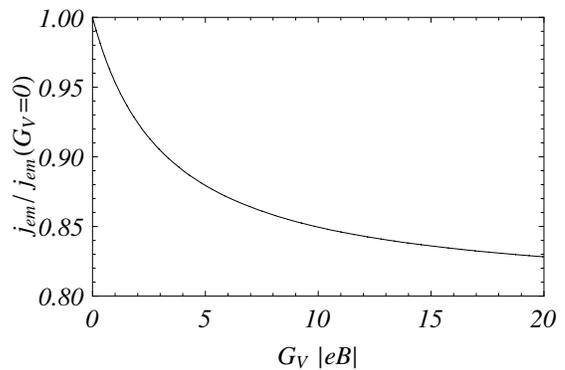}
 \caption{The dielectric correction coefficient as a function of
   $G_V|eB|$ in the two-flavor case.}
 \label{fig:kappa}
\end{figure}

\paragraph{Two-flavor case:}
The situation is drastically changed if there are multiple flavors
with different electric charges.  Now we consider a system having $u$
quarks with the electric charge $q_u=(2/3)e$ and $d$ quarks with
$q_d=(-1/3)e$.  The susceptibility becomes flavor dependent as
\begin{equation}
 \calC_R^f = \frac{N_c |q_f B|}{2\pi^2} \;,
\end{equation}
with which the current is expressed in the same way as
Eq.~\eqref{eq:jren} as follows;
\begin{equation}
 \bj = \frac{1}{\displaystyle 1+2G_V \sum_f \calC_R^f} \cdot
  N_c \sum_f \frac{\mu_5\,q_f \bB}{2\pi^2} \;.
\label{eq:jrenQCD}
\end{equation}
Up to this point the generalization is just straightforward.  In the
presence of non-degenerate flavors the electric current is no longer
proportional to $\bj$, but one has to compute $\bj_{\rm em}$ as the
flavor summation of $q_f$ times current contribution from each flavor
sector.  That is, the electric current density should be read from
\begin{equation}
 \bj_{\rm em} = N_c \sum_f \frac{q_f^2 \mu_5\bB}{2\pi^2}
  - 2 G_V \bj \sum_f q_f \calC_R^f \;.
\label{eq:jemQCD}
\end{equation}
After some calculations we find,
\begin{equation}
 \kappa = \biggl(1+\frac{12}{5\pi^2}G_V |eB|\biggr)
  \biggl(1+\frac{3}{\pi^2}G_V |eB|\biggr)^{-1} \,.
\label{eq:two}
\end{equation} 
This result~\eqref{eq:two} behaves different qualitatively from
Eq.~\eqref{eq:one}.  One can easily see that $\kappa$ asymptotically
approaches a finite number $4/5=0.8$ in the limit of $G_V\to\infty$
and $\kappa$ never goes to zero.  To show this clearly, we make a plot
for the above $\kappa$ as a function of $G_V|eB|$ in
Fig.~\ref{fig:kappa}.  It is obvious from the figure that $\kappa$
slowly decays to the asymptotic value $0.8$ and thus the dielectric
correction is only a minor effect in contrast to the one-flavor
situation.

\paragraph{Three-flavor case:}
It is interesting to think of the ideal case with three flavors.
Because the chiral magnetic current has the anomaly origin and is
independent of the quark masses, the three-flavor case might be
realistic if $\mu_5$ is large enough.  Let us imagine that the system
has $s$ quarks with $q_s=(-1/3)e$ in addition to $u$ and $d$ quarks.
Then the situation is totally changed again.  In view of
Eq.~\eqref{eq:jrenQCD}, $\bj$ is proportional to $\sum_f q_f$ which
becomes vanishing for the three-flavor case; $q_u+q_d+q_s=0$.  Because
$\bj$ is zero, there is no correction appearing at all, and thus
identically $\kappa=1$.  This result is intuitively understandable if
quark masses are degenerate;  the system is then automatically
electric-charge neutral.  Hence, there is no coupling between the
baryon current $\bj$ and the electric current $\bj_{\rm em}$ and thus
no back-reaction from their entanglement.  It is, however, non-trivial
that the conclusion of no correction for the three-flavor case holds
regardless of whether quark masses are degenerate or not.


Now let us plug concrete numbers in our final expression to see how
large/small the correction is in specific examples.  We first need to
determine a value of $G_V$.  From the discussions below
\eqref{eq:vector} we can postulate $G_V\sim g^2/M_g^2$ and let us
choose $g=2$ ($\alpha_s\sim0.3$) and $M_g=0.8\GeV$ (that is roughly a
half of the glueball mass) here.  Then we have a rough estimate as
$G_V\simeq 6.3\GeV^{-2}$.  We can make it sure that this is a
reasonable estimate from the empirically adopted $G_V$ in the NJL
model;  $G_V=0.2\sim0.5G_S$ where $G_S$ is the four-fermion coupling
in the scalar and pseudo-scalar channel and fixed as $G_S\simeq
9.2\GeV^{-2}$ to reproduce the pion mass and decay constant
\cite{Hatsuda:1994pi}.  This is not far from our estimate
$G_V\simeq 6.3\GeV^{-2}$.

In the heavy-ion collision it would be convenient to express the
magnetic field strength $|eB|$ in the unit of the pion mass squared
$m_\pi^2$ instead of gauss.  From the UrQMD simulation $|eB|$ is
evaluated as a few times $m_\pi^2$ for the RHIC energies
\cite{Skokov:2009qp}.  If we use the value $|eB|=m_\pi^2$, then, we
find $G_V|eB|=0.11$.  This is a small number and the dielectric
constant stays close to unity for any case of flavor number.
Therefore, fortunately, we can conclude that the dielectric correction
from back-reaction is only minor and practically negligible for
phenomenology.

Finally let us make a comment on a possible application of our result
to the lattice-QCD simulation.  The chiral magnetic effect has been
investigated in the lattice-QCD
simulation~\cite{Buividovich:2009wi,Abramczyk:2009gb} with extremely
strong magnetic fields.  For example, in
Ref.~\cite{Buividovich:2009wi}, the applied magnetic field can be as
strong as $|eB|\sim \GeV^2$, which is of order hundred in the unit of
$m_\pi^2$.  If we use $G_V\simeq 6.3\GeV^{-2}$ then $G_V|eB|\sim 6.3$
for $|eB|\sim 1\GeV^2$.  According to our expressions the one-flavor
system would lead to a substantial suppression factor
$\kappa\sim 0.34$.  Even in the two-flavor case the suppression is a
sizable effect; $\kappa\sim 0.87$.  Therefore, it should be possible
to confirm the existence of such dielectric corrections as discussed
here using the lattice-QCD simulation.

The lattice-QCD simulation opens an intriguing possibility that $G_V$
may be determined from $\kappa$.  In fact the determination of $G_V$
provides us with very useful information on the QCD phase diagram.
Especially it crucially depends on $G_V$ whether the chiral phase
transition can become of first order at finite density and whether the
QCD critical point can exist on the phase diagram.  Once a finite
baryon chemical potential is turned on, of course, the sign problem
hinders the simulation.  Nevertheless the two-color two-flavor
simulation is still feasible even at finite density, which may give
$G_V$ as a function of density, if the precise determination of chiral
magnetic current is possible from the lattice data.


In summary we computed the back-reaction coming from the vector
interaction, which should result in a dielectric correction on the
chiral magnetic current even at the mean-field level.  Our final
expressions show that the qualitative behavior of the correction
strongly depends on the relevant number of flavors in the system.  The
one-flavor case has a substantial suppression on the chiral magnetic
current due to screening effects, while the two-flavor case has only a
minor modification however strong the vector interaction is.  There is
no correction at all for the three-flavor case.  It should be possible
to quantify the correction in the lattice-QCD simulation, which in
turn would give useful information on the strength of the effective
vector interaction.


We thank I.~Shovkovy for inspiring discussions at the workshop on
``New Frontiers in QCD 2010'' at the Yukawa Institute for Theoretical
Physics.  M.~R.\ acknowledges discussions with H.~Abuki and
R.~Anglani.  The work of M.~R.\ is supported by JSPS under the
contract number P09028.  K.~F.\ is supported by Japanese MEXT grant
No.\ 20740134.


\end{document}